\documentclass{elsart3p}

\usepackage{graphicx}
\usepackage{amssymb}

\begin{document}

\begin{frontmatter}

%titles, authors and addresses

% use the thanksref command within \title, \author or \address for footnotes;
% use the corauthref command within \author for corresponding author footnotes;
% use the ead command for the email address,
% and the form \ead[url] for the home page:
% \title{Title\thanksref{label1}}
% \thanks[label1]{}
% \author{Name\corauthref{cor1}\thanksref{label2}}
% \ead{email address}
% \ead[url]{home page}
% \thanks[label2]{}
% \corauth[cor1]{}
% \address{Address\thanksref{label3}}
% \thanks[label3]{}

\title{Renormalization of thermal conductivity of disordered $d$-wave superconductors by impurity-induced local moments}
% use optional labels to link authors explicitly to addresses:
% \author[label1,label2]{}
% \address[label1]{}
% \address[label2]{}

\author[a]{Brian M. Andersen} and
\author[a,b]{P. J. Hirschfeld}

\address[a]{Department of Physics, University of Florida, Gainesville, Florida 32611-8440, USA}
\address[b]{Laboratoire de Physique des Solides, Universite Paris-Sud, 91405 Orsay, France}

\begin{abstract}
The low-temperature thermal conductivity $\kappa_0/T$ of $d$-wave
superconductors is generally thought to attain a ``universal" value
independent of disorder at sufficiently low temperatures, providing
an important measure of the magnitude of the gap slope near its
nodes. We discuss situations in which this inference can break down
because of competing order, and quasiparticle localization.
Specifically, we study an inhomogeneous BCS mean field model with
electronic correlations included via a Hartree approximation for the
Hubbard interaction, and show that the suppression of $\kappa_0/T$
by localization effects can be strongly enhanced by magnetic moment
formation around potential scatterers.
\end{abstract}

\end{frontmatter}

% main text
\section{Introduction}
\label{introduction} Thermal conductivity measurements at low
temperatures $T$ in the superconducting state have played an
important role in strengthening the case for a $d$-wave BCS-like
description of quasiparticles in  optimally doped cuprate
superconductors. They are bulk probes of the superconducting state,
unlike ARPES and STM, and can currently be performed at lower $T$
than microwave experiments.  One drawback is the need to separate
phonon and electron contributions, but in the cuprate
superconductors an asymptotic linear term, $\kappa_0/T\sim {\rm
const}$ which can be attributed solely to quasiparticles is almost
always present at the lowest temperatures.  After theoretical
predictions of the universality of low-$T$ quasiparticle transport
in nodal superconductors\cite{universal}, experimental confirmation
was obtained in a number of optimally doped
materials\cite{universalexpt}.  According to the theory, which
relies on the disorder-averaged self-consistent $T$-matrix
approximation (SCTMA), the low-$T$ thermal conductivity is given by
\begin{equation}
~~~~~~~~~~~~~\kappa_{00}={\kappa_0\over T} \simeq {k_B^2\over 3}
\left( {{v_F\over v_\Delta} + {v_\Delta\over v_F} }\right),
\label{eq1}
\end{equation}
where $v_\Delta$ is the gap slope at the node, and $v_F$ the Fermi
velocity.  This result has been shown to be insensitive to vertex
corrections due to anisotropic scattering\cite{DurstLee}. Since
$v_F$ is generally well-known, Eq.(\ref{eq1}) has been used to
extract the gap slope for a number of cuprates at lower doping as
well\cite{exptlunderdoped}, leading to the conclusion that
$v_\Delta$ increases with underdoping\cite{exptlunderdoped}.  This
conclusion is in apparent contradiction to  recent ARPES
experiments\cite{ARPES}, so it is worthwhile to examine physical
effects outside the framework of the SCTMA which could lead to a
suppression of $\kappa_0/T$ and thereby to the possible erroneous
conclusion based on Eq. (\ref{eq1}) that $v_\Delta$ was increasing.

There are several effects known to lead to a suppression of
$\kappa_0/T$. Localization effects were discussed in this context in
Ref. \cite{Atkinson}, and effects of bulk subdominant competing
orders have been shown to suppress $\kappa_0/T$ but do not
immediately eliminate it, despite the removal of the $d$-wave
nodes\cite{Gusynin}. Here, we investigate the effects on $\kappa(T)$
by local impurity-induced moments relevant e.g. to the underdoped
regime of La$_{2-x}$Sr$_x$CuO$_4$ (LSCO)\cite{bella}.

\section{Magnetic correlations.}
\label{correlations} The model used to study disordered $d$-wave
superconductors with magnetic correlations is:
\begin{eqnarray}\label{hamiltonian}
\hat{H}= &-& t \sum_{\langle ij \rangle\sigma}
\hat{c}_{i\sigma}^{\dagger}\hat{c}_{j\sigma} + \sum_{\langle
ij\rangle} \left( \Delta_{ij}
\hat{c}_{i\uparrow}^{\dagger}\hat{c}_{j\downarrow}^{\dagger} +
\mbox{H.c.} \right) \nonumber\\ &-& \sum_{i\sigma} \mu
\hat{n}_{i\sigma} + U\sum_{i\sigma} \hat{n}_{i\sigma} \langle
\hat{n}_{i\overline{\sigma}} \rangle + \sum_{i\sigma} V_i
\hat{n}_{i\sigma}.
\end{eqnarray}
Here, $\hat{c}_{i\sigma}^{\dagger}$ creates an electron of spin
$\sigma$ on the site $i$ on a two-dimensional lattice, $\mu$ is the
chemical potential and
$\hat{n}_{i\sigma}=\hat{c}_{i\sigma}^{\dagger}\hat{c}_{i\sigma}$ is
the particle number operator. In general, magnetic order induced by
$U$ will compete with the superconducting order $\Delta_{ij}$ and
lead to a bulk magnetic state above some critical $U_{c0}$. However,
even for $U_c <U<U_{c0}$, disorder $V_i \neq 0$ caused by point-like
impurities can induce localized $S=1/2$ states with staggered
magnetization. Here, $U_c$ is the critical $U$ necessary for local
moment formation, and will in general depend on the impurity
strength. In this paper we study impurity configurations with 2\%
randomly positioned point-like impurities of strength $V_i=35t$
which is in the unitary limit. We have solved Eq.
(\ref{hamiltonian}) selfconsistently on $40\times 40$ lattices with
hole doping $\delta=12\%$, and calculated the electronic $\kappa(T)$
as outlined in Ref. \cite{Atkinson}.

\begin{figure}[t]
\includegraphics[width=7.5cm,height=6.0cm]{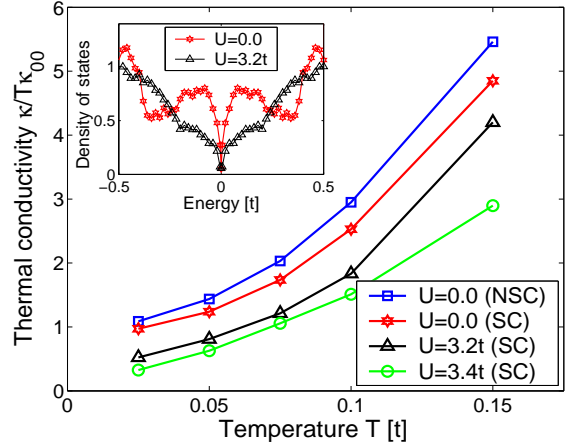}
\caption{$\kappa(T)/T\kappa_{00}$ versus $T$ for $2\%$ point-like
impurities with $U=0$ (non-selfconsistent (NSC) ($\square$),
selfconsistent (SC) ($\star$)), $U=3.2t$ ($\triangle$), and $U=3.4t$
($\circ$). The finite system size restricts us to study the region
$T/t \gtrsim 0.02$.\label{kappafig} Inset: Selfconsistent DOS for
the cases $U=0.0$ ($\star$) and $U=3.2t$ ($\triangle$).}
\end{figure}

In Fig. \ref{kappafig}, we show the thermal conductivity averaged
over 20 different random impurity configurations. For $U=0$ the
results agree well with those obtained previously for the dilute
impurity limit\cite{Atkinson}: $\kappa(T)/T = \kappa_{00} + \alpha
T^2$, and the spatial inhomogeneity of $\Delta_{ij}$ only slightly
modifies $\alpha$ as seen by comparing the SC and NSC results. For
the bulk system we have $\Delta_{ij}=0.4t$ on each link resulting in
the universal value $\kappa_{00}=1/3\left( {{v_F/v_\Delta} +
{v_\Delta/v_F} }\right)=0.967$, which is nicely reproduced by the
numerics. For $0 < U < U_c $, $\forall i: \langle
n_{i\uparrow}\rangle=\langle n_{i\downarrow}\rangle$ and $U$ enters
as a chemical potential shift which is compensated by a modified
$\mu$ (to get $\delta =12\%$) and consequently $\kappa(T)$ is
identical to the $U=0$ result. In the regime $U_c <U<U_{c0}$, local
moments are formed and the corresponding thermal conductivity is
also shown in Fig. \ref{kappafig}. As seen, when $U$ increases,
$\kappa(T)/T$ (and $\kappa_0/T$) is continuously suppressed, and
$\kappa_0/T$ will eventually vanish in the bulk insulating state
around $U \sim U_{c0}$. The origin of the suppressed thermal
conductivity can be traced to a reduction of the low-energy density
of states (DOS) by the local magnetic moments as shown in the inset
in Fig. \ref{kappafig}. Thus, impurity-driven local moment formation
in the underdoped regime may explain the doping dependence of
$\kappa_0/T$ measured in LSCO\cite{takeya}. In addition, since the
low-$T$ thermal conductivity is severely suppressed, measured values
of $\kappa_{00}$ and a naive use of the "universal" clean $d$-wave
result in Eq. (\ref{eq1}) would lead to an erroneous estimate the
superconducting gap.

%In conclusion we have shown that local moment formation can severely
%suppress the low-$T$ thermal conductivity, and would lead to an
%overestimate of the superconducting gap if it was naively extracted
%from the "universal" clean $d$-wave result in Eq. (\ref{eq1}).

{\it Acknowledgements} This work is supported by ONR grant
N00014-04-0060.

\end{document}